# Single crystal growth of $Cu_4(OH)_6BrF$ and universal behavior in quantum spin liquid candidates synthetic barlowite and herbertsmithite

C.M. Pasco[1,2], B.A. Trump[1,2], Thao T. Tran[1,2], Z.A. Kelly[1,2], C. Hoffmann[3], I. Heinmaa[4], R. Stern[4], and T.M. McQueen[1,2,5]

[1]Department of Chemistry, The Johns Hopkins University, Baltimore, MD 21218
[2]Institute for Quantum Matter, Department of Physics and Astronomy, The Johns Hopkins University, Baltimore, MD 21218
[3]Oak Ridge National Laboratory, Oak Ridge, TN 37831
[4]National Institute of Chemical Physics and Biophysics, Akadeemia tee 23, 12618 Tallinn, Estonia
[5]Department of Materials Science and Engineering, The Johns Hopkins University, Baltimore, MD 21218

**Abstract**

Synthetic barlowite, $Cu_4(OH)_6BrF$, has emerged as a new quantum spin liquid (QSL) host, containing kagomé layers of S=1/2 $Cu^{2+}$ ions separated by interlayer $Cu^{2+}$ ions. Similar to synthetic herbertsmithite, $ZnCu_3(OH)_6Cl_2$, it has been reported that $Zn^{2+}$ substitution for the interlayer $Cu^{2+}$ induces a QSL ground state. Here we report a scalable synthesis of single crystals of $Cu_4(OH)_6BrF$. Through x-ray, neutron, and electron diffraction measurements coupled with magic angle spinning $^{19}F$ and $^{1}H$ NMR spectroscopy, we resolve the previously reported positional disorder of the interlayer $Cu^{2+}$ ions and find that the structure is best described in the orthorhombic space group, *Cmcm*, with lattice parameters $a$ = 6.665(13) Å, $b$ = 11.521(2) Å, $c$ = 9.256(18) Å and an ordered arrangement of interlayer $Cu^{2+}$ ions. Infrared spectroscopy measurements of the O—H and F—H stretching frequencies demonstrate that the orthorhombic symmetry persists upon substitution of $Zn^{2+}$ for $Cu^{2+}$. Specific heat and magnetic susceptibility measurements of Zn-substituted barlowite, $Zn_xCu_{4-x}(OH)_6BrF$, reveal striking similarities with the behavior of $Zn_xCu_{4-x}(OH)_6Cl_2$. These parallels imply universal behavior of copper kagomé lattices even in the presence of small symmetry breaking distortions. Thus synthetic barlowite demonstrates universality of the physics of synthetic $Cu^{2+}$ kagomé minerals and furthers the development of real QSL states.

## I. INTRODUCTION

Quantum spin liquids (QSL) represent a new class of magnetic behavior characterized by long-range entanglement and have attracted considerable attention [1-4]. The suppression of classical long-range antiferromagnetic order by geometric frustration has been considered a promising route to the production of materials which could host a QSL state [5-7]. One of the most studied candidates for hosting this state is synthetic herbertsmithite (HBS), $ZnCu_3(OH)_6Cl_2$, a spin-1/2 kagomé antiferromagnet and the end member of the $Zn^{2+}$ substitution series of the $Zn_xCu_{4-x}(OH)_6Cl_2$ paratacamite family [8-18]. HBS possesses copper kagomé planes which are separated from each other by non-magnetic $Zn^{2+}$ ions in the interlayer. Recent work has shown that there are residual $Cu^{2+}$ ions (~15%) on the interlayer, complicating investigation of the low energy physics of the S=1/2 kagomé lattice [16-18].

Synthetic barlowite, $Cu_4(OH)_6BrF$, is another copper S = ½ kagomé compound also reported to harbor perfect $Cu^{2+}$ kagomé layers [19-21]. In contrast to synthetic herbertsmithite, the layers of synthetic barlowite directly overlap (AAA stacking) each other rather than being staggered (ABC stacking), resulting in a trigonal prismatic arrangement of hydroxide ions surrounding the interlayer copper/zinc site (instead of octahedral as in synthetic herbertsmithite). Theoretical work has indicated that the substitution of $Zn^{2+}$ or $Mg^{2+}$ on the interlayer sites of barlowite can be accomplished with significantly less disorder than seen in herbertsmithite [22, 23]. Recent experimental reports have shown that the substitution of $Zn^{2+}$ in the structure is possible and results in the possible formation of a spin liquid ground state [24, 25]. There are no reported methods for producing Zn substituted single crystals of synthetic barlowite suitable for single crystal neutron scattering studies.

Here we report the successful preparation of synthetic barlowite single crystals, with a technique that allows scaling to arbitrarily larger sizes. Using a combination of diffraction techniques complimented with magic angle spinning $^{19}F$ and $^{1}H$ NMR spectroscopy, we resolve the previously reported positional disorder of the interlayer $Cu^{2+}$ ions in $Cu_4(OH)_6BrF$ and find that the appropriate crystallographic symmetry of the room temperature phase is orthorhombic, *Cmcm*. Infrared spectroscopy measurements of the O—H and F—H stretching frequencies demonstrate that the orthorhombic symmetry persists upon substitution of $Zn^{2+}$ for $Cu^{2+}$ at least up to a $Zn^{2+}$ substitution of 0.46.

We also compare pure and Zn substituted barlowite, $Cu_{4-x}Zn_x(OH)_6BrF$, against the herbertsmithite compositional series, $Cu_{4-z}Zn_z(OH)_6Cl_2$, using specific heat and magnetic susceptibility measurements. Both series both contain near perfect copper kagomé layers which are isolated from each other by interlayer ions. The striking similarities in the behavior of these two series implies a universality of the evolution of a QSL state on these materials despite the presence of small symmetry breaking distortions seen in synthetic barlowite. Thus synthetic barlowite provides a useful comparison to probe the physics of synthetic $Cu^{2+}$ kagomé minerals and the development of potential QSL states in such materials.

## II. EXPERIMENTAL

### A. Polycrystalline Synthesis

Polycrystalline samples of $Cu_4(OH)_6BrF$ were grown through a hydrothermal reaction of 2 mmol copper carbonate basic (malachite $CuCO_3Cu(OH)_2$), with 4 mmol hydrobromic acid (HBr) and 4 mmol of ammonium fluoride ($NH_4F$) in 19 mL deionized water (DI $H_2O$) at 393K in a 23 mL Parr acid digestion vessel with PTFE liner for two to three days. At the end of the reaction the acid digestion vessels were removed from the furnace and allowed to cool in air to room temperature. It was also found that allowing the malachite and ammonium fluoride to react in the distilled water for 15 minutes before adding the HBr resulted in higher quality batches of polycrystalline sample, as did substitution of 3 mL of DI $H_2O$ with 3mL of ethanol (EtOH). Immediately following addition of HBr the vessels were sealed and the contents thoroughly mixed by shaking. Deuterated polycrystalline samples were grown in the same manner with the substitution of $D_2O$ for DI $H_2O$, DBr for HBr and singly deuterated ethanol, EtOD ($CH_3CH_2OD$), for EtOH.

$Zn^{2+}$ substitution was performed by including zinc carbonate basic (hydrozincite, $Zn_5(CO_3)_2(OH)_6$) at a 3:1.1 Cu:Zn ratio. The two samples measured using this method had 45.8(5)% and 46.3(5)% substitution of the interlayer copper site with zinc as determined by inductively coupled plasma mass spectrometry (ICP-MS). Additional hydrozincite in the reaction did not appear to improve the substitution of the interlayer site. Using 4 mmol of 1-Bromo-2,5-pyrrolidinedione in place of HBr with all other conditions the same as listed for the earlier $Zn^{2+}$ substitution reaction resulted in a 6% substitution of the interlayer copper site with zinc, using 1 mmol did not produce a usable sample.

### B. Single crystal growth

The synthesis of deuterated mm-scale single crystals was carried out through seeded, serial hydrothermal synthesis. There were two distinct stages to the growth of large single crystals: the first was used for cycles 1-14, the second was used for cycles 15-27.

#### *1. Cycles 1-14*

Initially 23 seed crystals, made using the polycrystalline method, with volumes around 0.004 $mm^3$ were placed in a 23 mL Parr acid digestion vessel with PTFE liner with 0.5 mmol $CuCO_3Cu(OH)_2$, 1 mmol $NH_4F$ and 1 mmol DBr in 19 mL $D_2O$ with a magnetic stir bar. The vessel was placed in a 393 K sand bath for 3 days with stirring before removing the vessel and allowing to cool to room temperature. After each growth cycle the crystals were mechanically cleaned of smaller crystals attached to their surfaces, then rinsed with water and ethanol to remove powder that remained on the surfaces. This process resulted in crystals up to 0.04 $mm^3$. Attempts at further growth using this method reduced the quality of new growth.

#### *2. Cycles 15-27*

The crystals were resurfaced by etching with DBr followed by mechanical removal of surface material. Five crystals were tracked through subsequent growths as shown in Fig. 1. The same procedure as was used for cycles 1-14 was used with the following changes: (1) no stirring, (2) oven heating, and (3) replacement of 2 mL of $D_2O$ with 2 mL of $CH_3CH_2OD$ (EtOD). Following each cycle the crystals were mechanically cleaned and then sonicated in a 1:1 mixture of distilled water and ethanol to remove any powder that remained on the surface. This process was repeated using the crystals obtained from each prior cycle.

### C. Characterization

The Cu:Zn ratio measured with an Agilent Inductively Coupled Plasma Mass Spectrometer with helium as the cell gas.

Powder x-ray diffraction (PXRD) data were collected at room temperature using a Bruker D8 Focus diffractometer with a LynxEye detector using Cu Kα radiation ($\lambda = 1.5424$ Å). Rietveld refinements on PXRD data were performed using Topas 4.2 (Bruker) [26]. Visualization was done in Vesta [27].

Single crystal X-ray diffraction (SXRD) data was collected at T = 110 K using the program CrysAlisPro (Version 1.171.36.32 Agilent Technologies, 2013) on a SuperNova diffractometer equipped with Atlas detector using graphite-monochromated Mo Kα (λ = 0.71073 Å). CrysAlisPro was also used to refine the cell dimensions and for data reduction. The temperature of the sample was controlled using the internal Oxford Instruments Cryojet. The structure was solved using SHELXS-97 and refined using SHELXL-97. [28]

Neutron powder diffraction data were collected using the BT-1 32 detector neutron powder diffractometer at the NCNR, NBSR. A Cu(311) monochromator with a 90° take-off angle, λ = 1.5397(2) Å, and in-pile collimation of 15 minutes of arc were used. Data were collected over the range of 3-168° *2-Theta* with a step size of 0.05°. The sample was loaded in a vanadium can sample container of length 50 mm and diameter 6 mm Data were collected under ambient conditions. Rietveld refinements were performed using GSAS [29] in EXPGUI [30].

Single crystal neutron diffraction (SND) data was collected on the TOPAZ beamline [31] at Oak Ridge National Laboratories Spallation Neutron Source at T = 100 K. Experiment planning [32, 33], visualization of data and indexing of the UB matrix was done using Mantid [34, 35, 36]. Data reduction and peak integration was done using ReduceSCD and anvred3. Refinement of single crystal neutron diffraction data was done using SHELXL-97 [28].

Synthetic barlowite samples were prepared for electron diffraction by grinding and dispersing the resultant power on a holey carbon coated Cu grid. Electron diffraction was performed using a Phillips CM300 atomic resolution transmission electron microscope with a field emission gun and an accelerating voltage of 300 kV operating in diffraction mode with a bottom mounted Orius SC1000A CCD detector and an exposure time of 3 seconds. Images were analyzed using ImageJ [37].

Magnetic susceptibility and specific heat data using the short pulse semiadiabatic method were collected on a Quantum Design Physical Properties Measurement System (PPMS).

IR spectra were collected using an iD5 diamond ATR Accessory on a Thermo Scientific Nicolet iS5 FT-IR Spectrometer using OMNIC software. Allowed IR-active modes were determined using the Bilbao crystallographic server. [38]

The magic-angle spinning (MAS) NMR spectra were recorded in a $\mu_0H$ = 4.7 T magnetic field around room temperature (RT) with spin echo pulse sequence on a Bruker AVANCE-II spectrometer using home built MAS probe for 1.8 mm od rotors. The spinning speed of the sample was 40 kHz (therefore the real temperature is probably slightly higher than RT), $^1$H resonance frequency was 200.06 MHz, with TMS reference. The $^{19}$F resonance frequency was 188.25 MHz, with a $CCl_3F$ reference. The computer fit was performed within Bruker Topspin program.

## III. RESULTS

Figure 1(a) shows the change in size of five single crystals of $Cu_4(OH)_6BrF$ as a function of the number of growth cycles. The volume of each crystal grows quadratically with the number of cycles, Fig. 1(b), implying that significantly larger crystals are possible by this technique. The power law, shown in the inset, suggests that the growth rate is surface area limited.

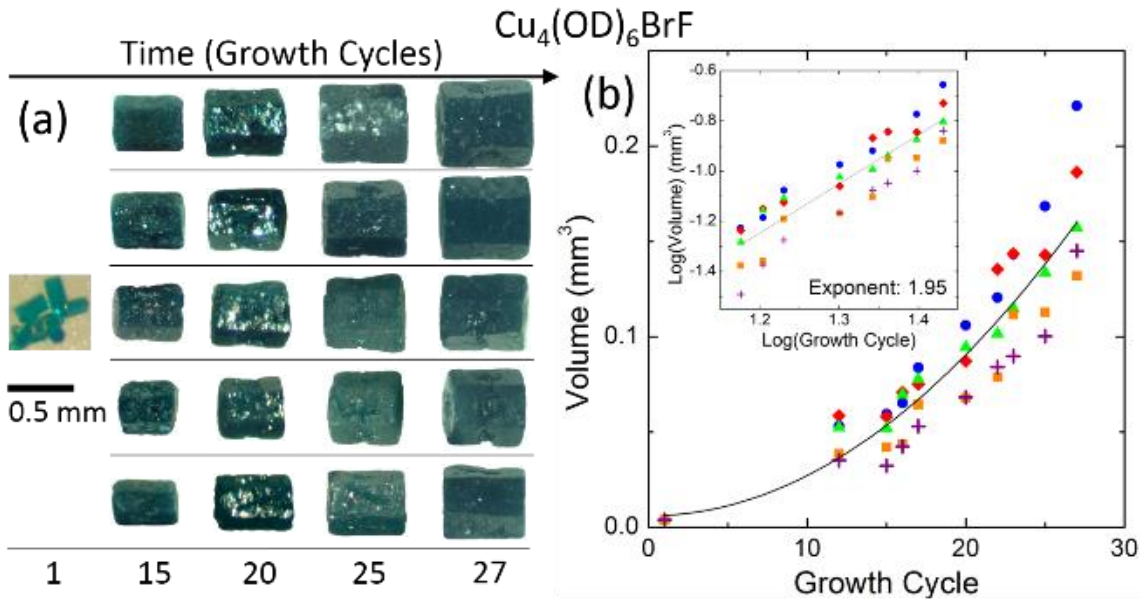

FIG. 1. (color online) (a) The five largest deuterated crystals followed over time. (b) Volumes of the crystals over time, the inset is a log-log plot from growth cycle 15 to 27 showing that the exponent of the growth rate is ~2, suggesting the growth rate is limited by the surface area of each crystal.

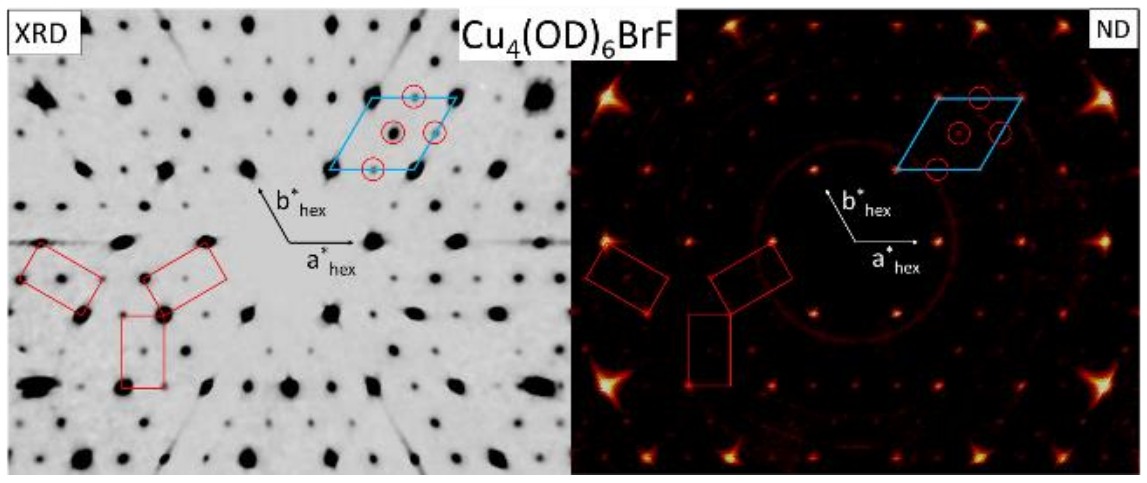

FIG. 2. (color online) (left) SXRD precession image indexed in *P6₃/mmc* in the (hk0) plane. (right) SND reciprocal space layer image indexed in *P6₃/mmc* in the (hk0) plane. The blue rhombus shows the unit cell reported for *P6₃/mmc* with peaks that it does not index circled in red. The red rectangles show the unit cells for the three *Cmcm* twins by pseudomerohedry. The SND reflection intensities are shown on a logarithmic scale, visible as well are powder rings from the sample mount.

PXRD measurements of ground single crystals and polycrystalline powders of $Cu_4(OH)_6BrF$ (not shown), and NPD measurements of polycrystalline $Cu_4(OD)_6BrF$ (not shown) are in excellent agreement with previous reports, and are indexable with the previously reported *$P6_3/mmc$* hexagonal unit cell [9].

SXRD data were collected on single crystals of barlowite. Weak reflections, not indexable by the reported *$P6_3/mmc$* structure, are clearly visible in the precession images, Fig. 2(a,b). In addition, our SXRD data shows that some, but not all, observed reflections exhibit a high degree of mosaicity. The high mosaicity does not affect spots at random but appears to affect a subset of the reflections significantly more than others, which is potentially indicative of twinning. SND also produces reflections which cannot be indexed in *P6₃/mmc* but are well correlated to those observed in the SXRD precession images. The highest symmetry unit cell compatible with the SXRD and SND data is an orthorhombic cell, space group *Cmcm*, with a = 6.67 Å, b = 11.52 Å, c = 9.26 Å. Inclusion of the appropriate twin law, ([-1 -1 0], [-0.5 -0.5 0], [0 0 -1] 3), allows this cell to index all observed reflections, with low mosaicity spots indexed by a single twin and high mosaicity spots indexed by multiple twins, Fig. 2(a,b). This particular type of twinning by pseudomerohedry, with a *Cmcm* structure masquerading as the higher symmetry space group, *P6₃/mmc*, is well-known in related materials [39, 40]. Recent powder neutron studies of the low temperature structure of synthetic barlowite have also indicated an orthorhombic space group, proposed to be *Pnma*, from the evolution of magnetic peaks below T = 15 K, though we find *Pnma* unable to describe the room temperature structure [25].

There is a second way that the additional reflections observed in the SXRD data can be fit, an enlarged hexagonal unit cell with lower symmetry, such as *P6₃/m*. This was tested for using electron diffraction (ED) shown in Fig. 3. ED measures a much smaller region of the sample than other diffraction methods and thus can resolve diffraction patterns from individual twin domains. The collected ED patterns are well indexed by the *Cmcm* cell and conclusively rule out a larger hexagonal cell with *P6₃/m* symmetry as it cannot generate the observed diffraction data. Synthetic barlowite is susceptible to degradation in an electron beam and so more detailed analysis could not be made, although it should be noted that *P6₃/mmc* does not fit the observed pattern of intensities.

The final structure of synthetic barlowite, from refinement of the SND data in the space group *Cmcm*, is given in Table 1. This model adequately describes the data from all diffraction measurements. For *Cmcm* the overall refinement parameter was $R(F_0) = 0.0603$ which is a significant improvement over the $R(F_0) = 0.0824$ obtained for the disordered *P6₃/mmc*, further supporting the choice of *Cmcm*. Lower symmetry subgroups of *Cmcm* were also examined but did not improve the refinement (e.g. *P2₁/m* with $R(F_0) = 0.0604$ despite adding an additional 48 refineable parameters).

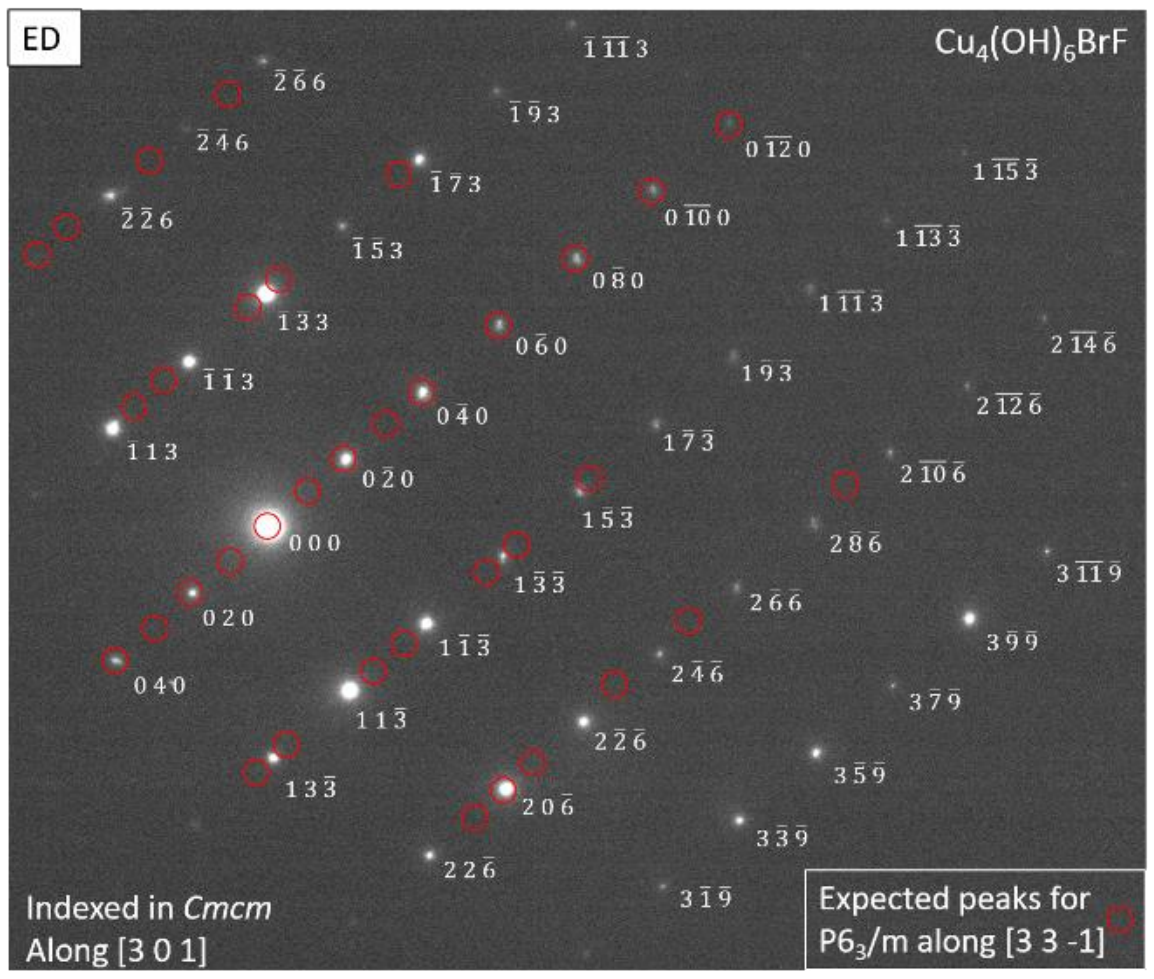

FIG. 3. (Color online) Electron diffraction pattern indexed to the *Cmcm* cell in white. Doubling the *a* and *b* axes of the reported *P6₃/mmc* and dropping the symmetry to *P6₃/m* cell results in expected bragg reflections (red circles) in poor agreement with the observed pattern.

TABLE I. Refinement of SND data to twinned *Cmcm*. Anisotropic thermal parameters are provided as a crystallographic information file in the SI.

| | | | |
|---|---|---|---|
| Space group | *Cmcm* | $\rho_{calc}$(g/cm$^3$) | 4.31 |
| *a* (Å) | 6.665(13) | Z | 4 |
| *b* (Å) | 11.521(2) | Total refl. | 5939 |
| *c* (Å) | 9.256(18) | Unique Refl. | 665 |
| α (°) | 90 | Parameters | 55 |
| β (°) | 90 | $R_{int}$ | 0.1389 |
| γ (°) | 90 | GooF | 1.225 |
| Temperature | 100 K | $R(F_0)$ | 0.0603 |
| Crystal Size (μm) | 628x628x670 | $R_w(F_o)^2$ | 0.1327 |

| Atom | Site | *x* | *y* | *z* | $U_{iso}$ (Å$^2$) |
|---|---|---|---|---|---|
| Cu(1) | 4*b* | ½ | 0 | 0 | 0.0114(16) |
| Cu(2) | 8*d* | ¼ | ¼ | ½ | 0.0124(9) |
| Cu(3) | 4*c* | 1 | 0.3713(4) | ¾ | 0.0077(7) |
| Br | 4c | 0 | 0.3327(8) | ¼ | 0.0143(11) |
| F | 4*c* | 0 | 0.9923(15) | ¼ | 0.0127(12) |
| O(1) | 8f | ½ | 0.2979(8) | 0.4125(9) | 0.0099(13) |
| O(2) | 16h | 0.8045(9) | 0.4010(5) | 0.9052(5) | 0.0107(7) |
| D(1) | 8f | ½ | 0.3762(7) | 0.3696(13) | 0.0258(19) |
| D(2) | 16h | 0.6865(1) | 0.4374(7) | 0.8656(5) | 0.0244(10) |

The *Cmcm* structural model is also in agreement with local measurements from magic-angle spinning (MAS) $^1$H and $^{19}$F NMR performed on a powder sample of synthetic barlowite. MAS suppresses line broadening from crystalline anisotropy and allows for greatly enhanced sensitivity to local structural orders. A typical MAS NMR spectrum consists of a resonance at an isotropic chemical shift value and several spinning sidebands at frequencies separated from the main line by a multiple of sample spinning frequency. One chemical shift should be observed for each crystallographically independent atomic position. In the highly symmetry constrained hexagonal *P6₃/mmc* space group only a single isomer shift should be expected for $^1$H. The $^1$H MAS-NMR spectrum of $Cu_4OH)_6BrF$, Fig. 4(a), clearly shows two distinct resonances at -111 ppm and -134 ppm, in addition to the spinning sidebands and a signal from probe background, making it inconsistent with *P6₃/mmc*.

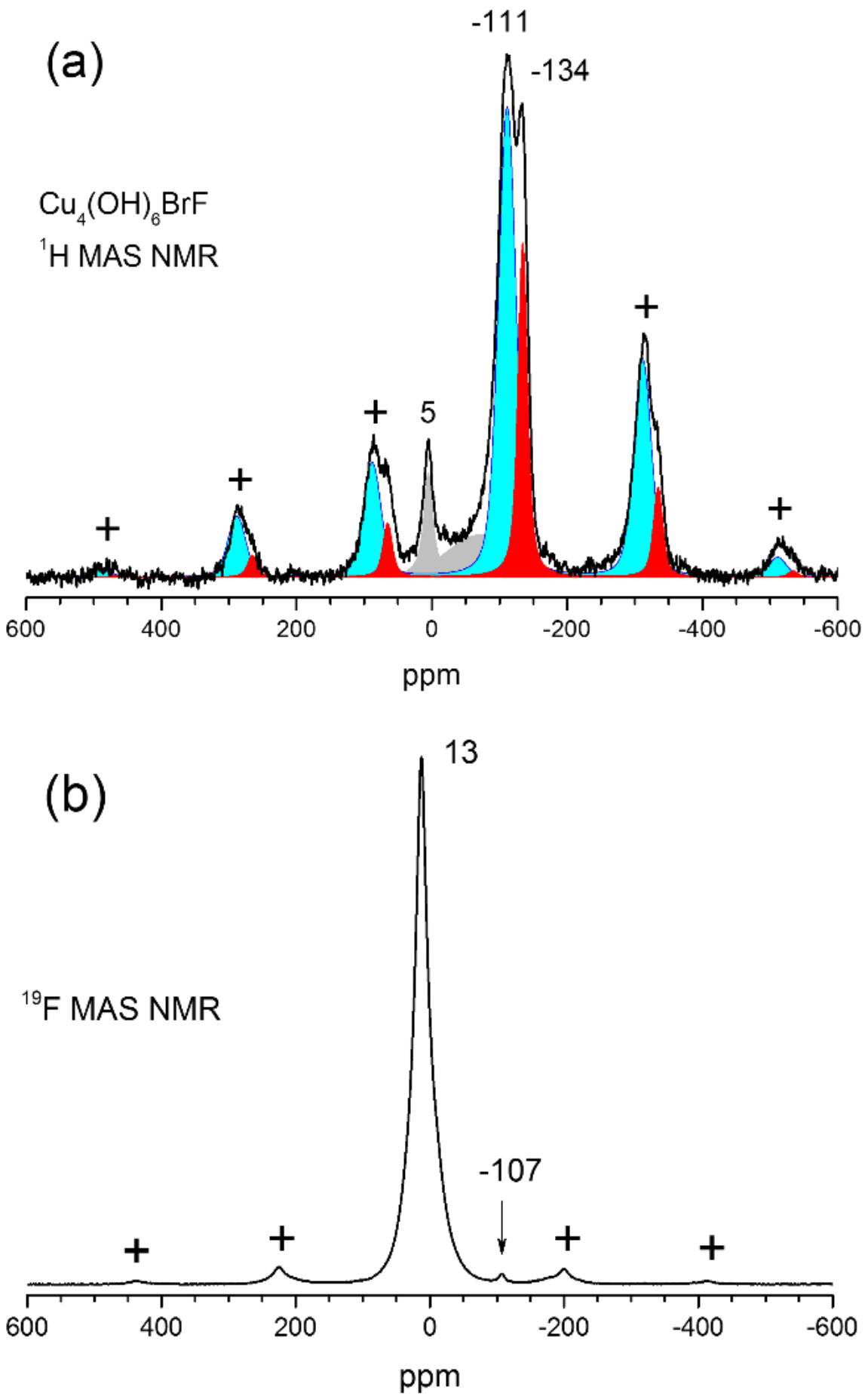

FIG. 4. (color online) The x axis shows the measured chemical shift and the y axis is an arbitrary intensity. (a) $^1H$ MAS NMR spectrum of $Cu_4(OH)_6BrF$ powder sample showing two hydrogen resonance lines at -111 (cyan) and -134 ppm (red) together with their spinning sidebands noted by +. The intensity ratio of the two resonances is 70:30. The peak at 5 ppm and a broad line noted in gray represent a proton background of the probe. (b) $^{19}F$ MAS NMR spectrum of the same sample shows a single fluorine line with a small chemical shift of 13 ppm and a tiny resonance at -107 ppm of unknown origin.

The intensity ratio 70:30 of the two resonances, as found from the computer fit, is close to the ratio 4:2 expected for *Cmcm* structure. Distribution of the spinning sideband intensities yields considerable anisotropy of the chemical shift (here magnetic hyperfine shift) tensor. The principal values of the tensors are estimated as {310, -300, -345} ppm and {210, -295, -315} ppm for the two lines. The $^{19}F$ MAS-NMR spectrum, Fig. 4(b) shows a single fluorine resonance in agreement with the *Cmcm* structure. In both $^1H$ and $^{19}F$ measurements, there is broadening beyond the expected instrumental resolution. In magnetic materials this is caused usually by the magnetic susceptibility of the powder particles [41]. Further the observed $^1H$ chemical shifts agree with the structure from SND: D1 (8 per cell) has closer $Cu^{2+}$ and $F^-$ ions than D2 (16 per cell) and should therefore have a larger hyperfine shift. Thus, the stronger line at -111 ppm should belong to D2 sites and the weaker line at -133 ppm belongs to D1, as expected from the ratios of intensities.

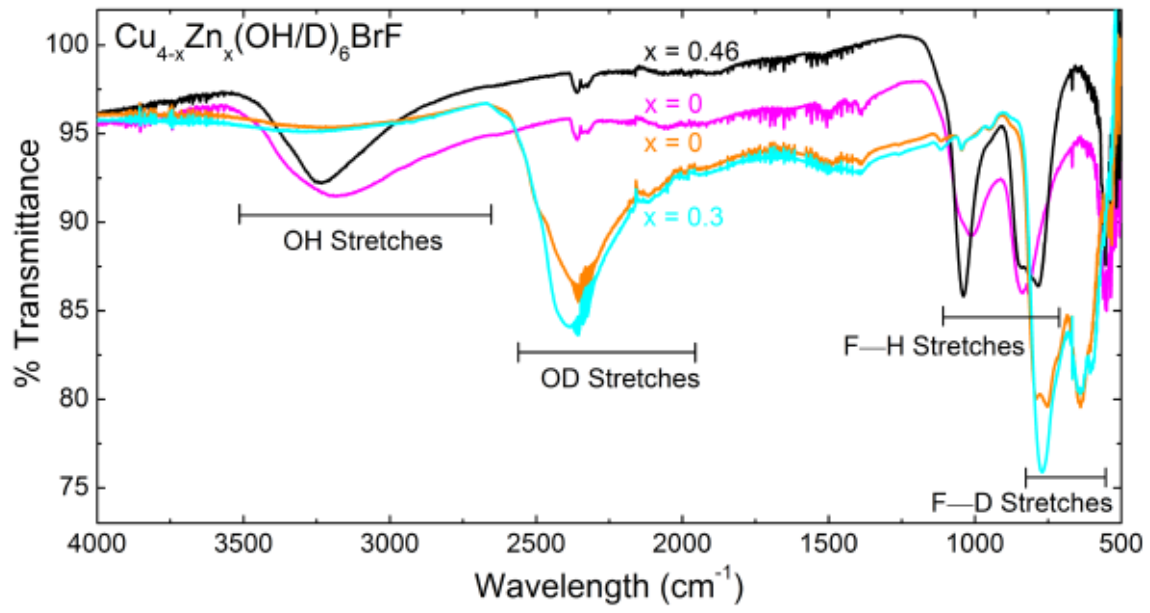

FIG. 5. (color online) ATR-IR measurement reported in transmittance mode. Partial $Zn^{2+}$ substitution is marked by a sharpening and a shift to higher wavenumber for the OH stretches and a shift to lower wavenumber for the F-H stretches.

TABLE II. locations of transmittance minima of ATR-IR peaks

| Sample | O-$^n$H peak | F—$^n$H peaks | | |
|---|---|---|---|---|
| x=0, $^1H$ | 3182 $cm^{-1}$ | 1044 $cm^{-1}$ | 1012 $cm^{-1}$ | 839 $cm^{-1}$ |
| x=0.4, $^1H$ | 3236 $cm^{-1}$ | 1039 $cm^{-1}$ | 840 $cm^{-1}$ | 781 $cm^{-1}$ |
| x=0, $^2H$ | 2350 $cm^{-1}$ | 789 $cm^{-1}$ | 752 $cm^{-1}$ | 639 $cm^{-1}$ |
| x=0.3, $^2H$ | 2382 $cm^{-1}$ | 770 $cm^{-1}$ | 639 $cm^{-1}$ | 601 $cm^{-1}$ |

We found that measurement of Γ-point phonon modes by attenuated total reflection infrared spectroscopy (ATR-IR) was a versatile measurement technique to probe the symmetry of synthetic barlowite when doped with Zn, i.e. $Zn_xCu_{4-x}(OH)_6BrF$. Representative data for hydrogenated and deuterated samples at x = 0, 0.3, and 0.4 are shown in Fig. 5. Absorption bands for O—H(D) and F—H(D) are clearly resolved at 3200 $cm^{-1}$ (2400 $cm^{-1}$) and 1000 $cm^{-1}$ (700 $cm^{-1}$) respectively, and tabulated in Table 2.

In *P6₃/mmc* there should only be two modes observable in IR related to the F—H(D) stretches, $1E_{1u}$ + $1A_{2u}$, however three distinct absorption peaks are clearly resolved. The observation of three IR active modes related to the F—H(D) stretches is expected for the *Cmcm* structure, as the reduction in symmetry results in $B_{1u}$ + $B_{2u}$ + $B_{3u}$ modes for the F—H(D) stretches, consistent with the observed data. The clustering of two of these three modes close together is also expected, as the *Cmcm* structure is very close to having the higher P63/mmc symmetry. Zinc substitution is found to slightly reduce the stretching frequencies, as expected, but does not change the number of observed modes. This implies that up to x = 0.5, the orthorhombic symmetry is maintained at room temperature. The significant broadening in the O—H(D) region precludes detailed analysis, but the overall shape is consistent with expectations for the *Cmcm* structure.

To explore similarities and differences between these two quantum spin liquid candidates, DC magnetization measurements of Zn substituted barlowite series were carried out from T = 2-300K. A comparison of the susceptibility, estimated as $\chi \approx M/H$, to literature data on the synthetic herbertsmithite family, measured at $\mu_0 H = 0.01$ T is shown in Fig. 6(a). In synthetic barlowite, $Zn^{2+}$ substitution suppresses both the magnitude of the ferromagnetic response at low temperature, and the temperature of the transition. This is similar to the behavior of $Zn_zCu_{4-z}(OH)_6Cl_2$, although the suppression of the ordering temperature is more pronounced in the case of $Zn_xCu_{4-x}(OH)_6BrF$.

Curie-Weiss analysis of the high temperature regime of each $Zn_xCu_{4-x}(OH)_6BrF$ sample is shown in Fig. 6(b), with corresponding parameters in Table 3. Increased Zn substitution lowers the temperature at which deviations from the Curie-Weiss law are observed, and the observed Weiss temperature becomes more negative. The Weiss temperatures of the barlowite series are a factor of ~2 smaller than that of herbertsmithite, suggesting ~two times weaker in-plane magnetic interactions.

The normalized plot of Fig. 6(c) provides a more meaningful comparison between synthetic barlowite and synthetic herbertsmithite. For $T/|\theta| > 1$, all curves follow linear behavior, as expected. Deviations from Curie-Weiss behavior are observed below $T/|\theta| = 1$ for all samples except the endmembers of both the herbertsmithite and barlowite $Zn^{2+}$ substitution series in which deviation occurs around $T/|\theta| < 0.3$. This suggests that interactions between kagomé and interlayer copper ions are crucial in governing the observed magnetic behavior.

Table III. Values from the Curie Weiss analysis. Linearized between 150 and 300K with the exception of z = 0 sample which was linearized between 200 K and 300 K. Values for z = 0, x = 0.92 done from reanalysis of data in refs. [42] and [24]

| Sample | C | θ (K) | $\chi_0$ |
|---|---|---|---|
| x = 0 | 0.34 | -78 | 1.5E-4 |
| x = 0.06 | 0.31 | -82 | 1.1E-4 |
| x = 0.46 | 0.32 | -89 | 9E-5 |
| x = 0.92 | 0.42 | -153 | 9E-5 |
| z = 0 | 0.43 | -155 | 3E-5 |
| z = 1 | 0.42 | -200 | -6E-5 |

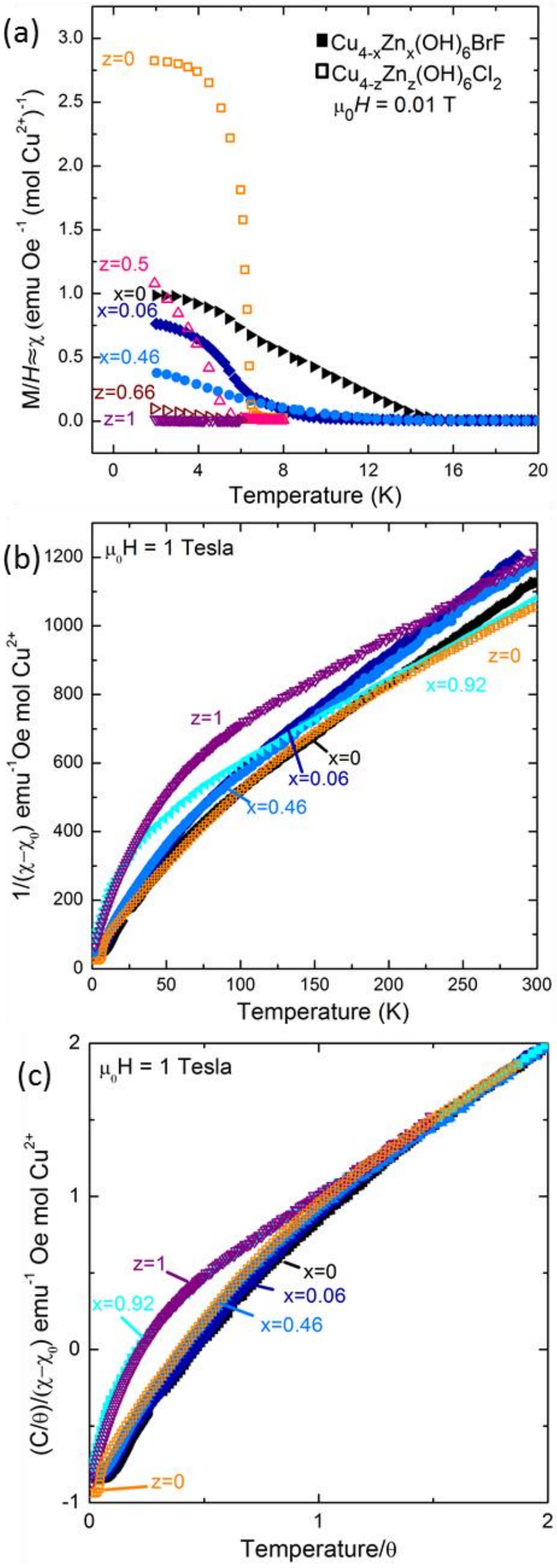

FIG. 6. (Color online) Symbols and colors are consistent for (a) through (c). Solid symbols represent $Cu_{4-x}Zn_x(OH)_6BrF$ data going from black to light blue with increasing zinc content. Hollow symbols represent $Cu_{4-z}Zn_z(OH)_6Cl_2$ data going from orange to purple with increasing zinc content. (a) $\chi_{mol}$ for $Zn^{2+}$ substitution series of barlowite compared to the herbertsmithite series. (b) Curie-Weiss analysis of all data sets. (c) Inverse susceptibility normalized for $C/|\theta|$ versus $T/|\theta|$, highlighting universal behavior at low x,z and high x,z. Data for z = 0, 0.5 0.66 from ref. [8, 43], data for x = 0.92 from ref. [24]. For (a), data for z = 1 from [8].

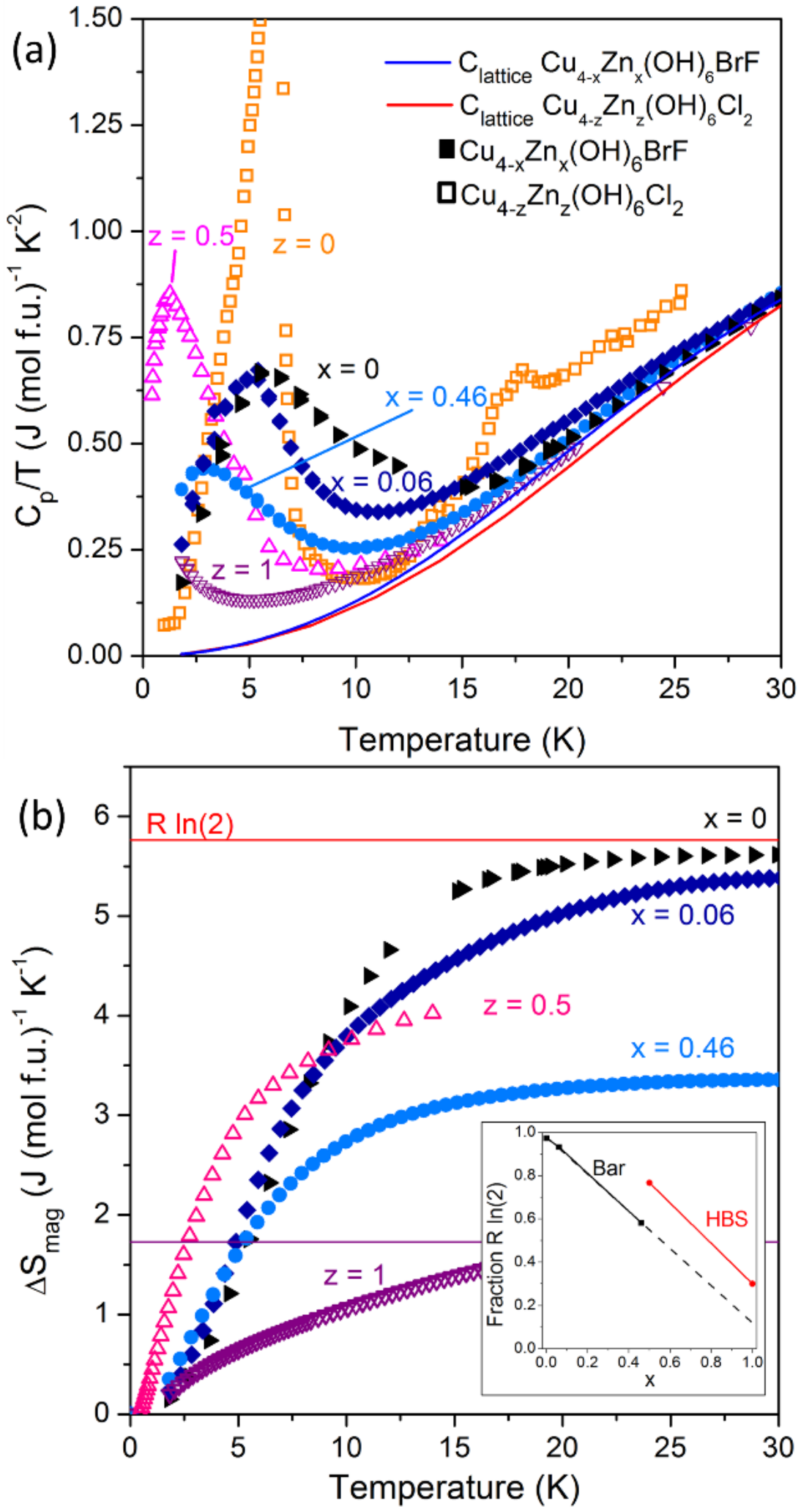

FIG. 7. (color online) Symbols and colors are consistent for (a) and (b). Solid symbols represent $Cu_{4-x}Zn_x(OH)_6BrF$ data going from black to light blue with increasing zinc content. Hollow symbols represent $Cu_{4-z}Zn_z(OH)_6Cl_2$ data going from orange to purple with increasing zinc content. (a) Heat capacity divided by temperature as a function of temperature under zero field for the barlowite substitution series compared to members of the herbertsmithite series. barlowite shows a similar shift in its antiferromagnetic transition towards lower temperature as seen in the herbertsmithite series. Approximated lattice contributions are shown in blue for barlowite and red for herbertsmithite. The data for the z = 0 series is cut off peaking at 3.5 J mol$^{-1}$ K$^{-2}$. Data for z = 0 from ref. [42] and data for z = 0.5 from ref. [11] (b) Recovered magnetic entropy as a function of temperature is shown for both series, with R ln(2) shown at the top and a line showing the total recovered entropy for z = 1 where it is cut off by the inset. The inset shows the recovered entropy as a fraction of R ln(2) with the barlowite series in black and the herbertsmithite series in red compared to the measured molar fraction of zinc with the dashed line projected from the linear behavior at the substitution levels measured, the projected remnant magnetic entropy at full substitution is significantly less than seen in herbertsmithite.

Zero field heat capacity (HC) measurements for T = 1.8 K to 300K further highlight the similarities between the two material families. As non-magnetic structural analogs are unknown, the high temperature data, T = 25-300K, was fit to a three component Debye model to estimate the lattice contributions. The lattice contributions estimated in this manner are shown as solid lines in Fig. 7(a). There are numerous similarities and differences between the two series.

The only material to show a sharp lambda anomaly, indicative of a proper phase transition, is $Cu_4(OH)_6Cl_2$. In contrast, $Cu_4(OH)_6BrF$ only shows a broad hump below the T = 15 K transition seen by magnetization measurements. This likely reflects the high degree of twinning found in synthetic barlowite, and matches the behavior of $Zn_zCu_{4-z}(OH)_6Cl_2$ for z>0.33 [8]. Similarly broad phase transitions have been observed in related materials with disorder including $FeSc_2S_4$ [43] and $Yb_2Ti_2O_7$ [44].

In both materials families, the observed low temperature peak in $C_p/T$ shifts to lower temperatures and decreases in magnitude as $Zn^{2+}$ substitution increases. Subtraction of the approximated lattice contribution and integration of $C_p/T$ vs. T yields estimates for the changes in magnetic entropy at low temperature in each sample shown in Fig. 7(b). $Cu_4(OH)_6BrF$ recovers very near R ln(2) in magnetic entropy. This suggests that only ¼ of the Cu ions lose their entropy at the T = 15 K transition. It is thus alluring to attribute the ordering in synthetic barlowite as arising from the interlayer $Cu^{2+}$ (1/4 of all Cu ions). As $Zn^{2+}$ is substituted, the recovered magnetic entropy is suppressed in a manner very similar to what is seen in synthetic herbertsmithite. The non-zero low temperature magnetic entropy in z = 1 $Zn_zCu_{4-z}(OH)_6Cl_2$ is attributed to the residual $Cu^{2+}$ known to exist on the interlayer site [17]. Promisingly for the barlowite series the recovered entropy far more closely matches the measured $Zn^{2+}$ substitution than seen in the herbertsmithite series as shown by the inset in Fig. 7(b), suggesting more selective substitution at the interlayer position.

## IV. DISCUSSION

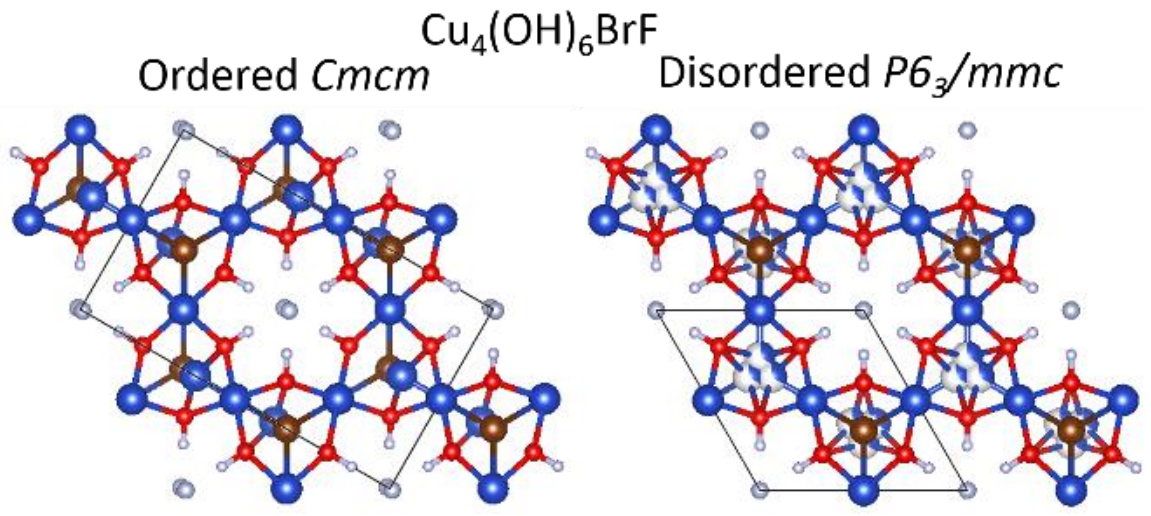

FIG. 8. Shown along the c-axis of the lattice is (left) the ordered *Cmcm* structure, and (right) the previously reported disordered *P6₃/mmc* structure [19]. The *Cmcm* structure results in local ordering of the interlayer coppers (blue) and distortion of the fluorine (white, center of channels) towards one pair of hydrogens (white, attached to oxygen atoms, red).

Figure 8 shows a comparison of the orthorhombic *Cmcm* structure of $Cu_4(OH)_6BrF$ to the previously reported, disordered, $P6_3/mmc$, model. Both structures share virtually identical kagomé layers, and in both, the trigonal prismatic coordination of the interlayer ions is highly unfavorable for either $Cu^{2+}$ or $Zn^{2+}$, requiring a displacement of the interlayer ion from its "ideal" position within the center of the trigonal prism into a distorted octahedral position. In $P6_3/mmc$ there are three equivalent distorted octahedral sites available to each interlayer copper and it is disordered across them.

The impact of the *Cmcm* model is very similar, but symmetry constraints no longer enforce the degeneracy of these three sites. The result is the ordering of the interlayer coppers into distinct positions. This ordering is coupled to a change in bonding of the fluorine atoms that sit in the hexagonal channels. In the previously reported structure, the $F^-$ ion sits in the center of a cage of $OH^-$ ions, with no direct bonding to any neighbors. This chemical coordination is extremely rare, having previously only been observed in a highly constrained zeolite [45]. If instead of being constrained to the center of the channel, the $F^-$ ions displace towards a pair of $OH^-$ ions, it can form optimal H—F hydrogen bonds. This is exactly what happens in the *Cmcm* structure, with corresponding motions of the interlayer $Cu^{2+}$ ions to accommodate the change in OH positions.

Remarkably, despite this deviation from perfect hexagonal symmetry, the physics of the kagomé layers appears mostly undisturbed, as seen in the comparison to the behavior of the herbertsmithite series. Since the gapped quantum spin liquid should be a stable phase on the kagomé lattice, it is not surprising that it should be stable against a small structural change.

While more work will be needed to fully understand the $Cu_{4-x}Zn_x(OH)_6BrF$ substitution series, that the recovered magnetic entropy more closes matches the measured level of zinc substitution than what is observed in the herbertsmithite series suggests that synthetic barlowite may allow the QSL state to be probed with fewer complications from remnant magnetic impurities. Our work demonstrates a method which makes possible the synthesis of large single crystals of Zn substituted barlowite sufficient to allow inelastic neutron scattering studies of the QSL state in this promising material.

## ACKNOWLEDGEMENTS

We acknowledge stimulating discussions with Collin Broholm. We thank Maxime Siegler for assistance with the single crystal x-ray diffraction measurements, Craig Brown for assistance with the powder neutron diffraction measurements, and R. Campbell-Kelly at the Westinghouse Electric Company Columbia Fuel Fabrication Facility for performing ICP-MS measurements. Work at the Institute for Quantum Matter was supported by the U.S. Department of Energy, Office of Basic Energy Sciences, Division of Material Sciences and Engineering under grant DEFG02-08ER46544. A portion of this research used resources at the Spallation Neutron Source, a DOE Office of Science User Facility operated by the Oak Ridge National Laboratory. We acknowledge the support of the National Institute of Standards and Technology, U. S. Department of Commerce, in providing the powder neutron diffraction facilities used in this work. Certain commercial equipment, instruments, or materials are identified in this paper to foster understanding. Such identification does not imply recommendation or endorsement by the National Institute of Standards and Technology, nor does it imply that the materials or equipment identified are necessarily the best available for the purpose. This research used resources of the Advanced Photon Source, a U.S. Department of Energy (DOE) Office of Science User Facility operated for the DOE Office of Science by Argonne National Laboratory under Contract No. DE-AC02-06CH11357. TMM acknowledges support of the David and Lucile Packard Foundation. I.H. and R.S. are supported by the Estonian Research Agency grants IUT23-7 and PRG4, and the European Regional Development Fund project TK134.